\documentclass[sigconf,nonacm]{acmart}

\AtBeginDocument{%
  }

\usepackage{tikz}
\usepackage{amsmath}
\usepackage{amsthm}
\usepackage{subcaption}
\usepackage{pifont}
\usepackage{makecell}
\usepackage{booktabs, multirow}
\theoremstyle{plain}
 
\usepackage{wrapfig}
\usepackage{balance} 
\usepackage{arydshln} 
\usepackage{algorithm}
\usepackage{algpseudocode}

\usepackage{todonotes}

\usepackage{mathtools, nccmath}
\DeclarePairedDelimiter{\nint}\lfloor\rceil
\DeclarePairedDelimiter{\abs}\lvert\rvert

\def\approxprop{%
  \def\p{%
    \setbox0=\vbox{\hbox{$\propto$}}%
    \ht0=0.6ex \box0 }%
  \def\s{%
    \vbox{\hbox{$\sim$}}%
  }%
  \mathrel{\raisebox{0.7ex}{%
      \mbox{$\underset{\s}{\p}$}%
    }}%
}

\newif\ifcomm

\ifcomm
\else
\commfalse
\fi
\ifcomm
\newcommand\jl[1]{\textcolor{violet}{JL: #1}}
\newcommand\MM[1]{\textcolor{orange}{\textbf{MM: #1}}}
\newcommand\ran[1]{\textcolor{blue}{Ran: #1}}
\newcommand{\cpq}[1]{{\color{red}[CPQ: #1]}}
\newcommand{\ga}[1]{{\color{red}[GA: #1]}}
\newcommand{\al}[1]{{\color{red}[AL: #1]}}
\else
\newcommand\jl[1]{}
\newcommand\MM[1]{}
\newcommand\ran[1]{}
\newcommand\cpq[1]{}
\newcommand\ga[1]{}
\newcommand\al[1]{}
\fi

\newif\ifblind

\renewcommand\footnotetextcopyrightpermission[1]{} 
\setcopyright{none}
\acmConference[Preprint]{}{}{}

\begin{document}

    \date{}
    
    \title{Sketch Disaggregation Across Time and Space}

\ifblind
    \author{Authors Anonymous}
\else
    \author{Jonatan Langlet}
    \affiliation{%
      \institution{KTH Royal Institute of Technology}
      \city{Stockholm}
      \country{Sweden}
    }
    \email{jlanglet@kth.se}

    \author{Peiqing Chen}
    \affiliation{%
      \institution{University of Maryland}
      \city{College Park}
      \state{MD}
      \country{United States}
    }
    \email{pqchen99@umd.edu}

    \author{Michael Mitzenmacher}
    \affiliation{%
      \institution{Harvard University}
      \city{Cambridge}
      \state{MA}
      \country{United States}
    }
    \email{michaelm@eecs.harvard.edu}

    \author{Ran Ben Basat}
    \affiliation{%
      \institution{University College London}
      \city{London}
      \country{United Kingdom}
    }
    \email{r.benbasat@ucl.ac.uk}

    \author{Zaoxing Liu}
    \affiliation{%
      \institution{University of Maryland}
      \city{College Park}
      \state{MD}
      \country{United States}
    }
    \email{zaoxing@umd.edu}

    \author{Gianni Antichi}
    \affiliation{%
      \institution{Politecnico di Milano}
      \city{Milan}
      \country{Italy}
    }
    \email{gianni.antichi@polimi.it}
    \renewcommand{\shortauthors}{Langlet et al.}  
\fi

    \begin{abstract}

Streaming analytics are essential in a large range of applications, including databases, networking, and machine learning. To optimize performance, practitioners are increasingly offloading such analytics to network nodes such as switches. However, resources such as fast SRAM memory available at switches are limited, not uniform, and may serve other functionalities as well (e.g., firewall). Moreover, resource availability can also change over time due to the dynamic demands of in-network applications.

In this paper, we propose a new approach to disaggregating data structures over time and space, leveraging any residual resource available at network nodes. 
We focus on sketches, which are fundamental for summarizing data for streaming analytics while providing beneficial space-accuracy tradeoffs. Our idea is to break sketches into multiple `fragments' that are placed at different network nodes.
The fragments cover different time periods and are of varying sizes, and are combined to form a network-wide view of the underlying traffic. 
We apply our solution to three popular sketches (namely, Count Sketch, Count-Min Sketch, and UnivMon) and demonstrate we can achieve approximately a $75\%$ memory size reduction for the same error for many queries, or a near order-of-magnitude error reduction if memory is kept unchanged.


    \end{abstract}
    
    \maketitle

    \section{Introduction}
\sloppy
Streaming analytics tools for tracking and analyzing system behavior, where requests are made at run-time, are important for many applications in databases~\cite{spark,google-file-system,distcache} and networking~\cite{sonata,elasticsketch,OpenSketch}.  
For example, in databases, one may wish to know the entries that are most frequently being accessed over a time window~\cite{distcache}; similarly, in networking, one may wish to find the heaviest flow (connection) in the network over a time period~\cite{basat2018volumetric}.  

\begin{figure}[t]
    \centering
    \includegraphics[width=1.0\linewidth]{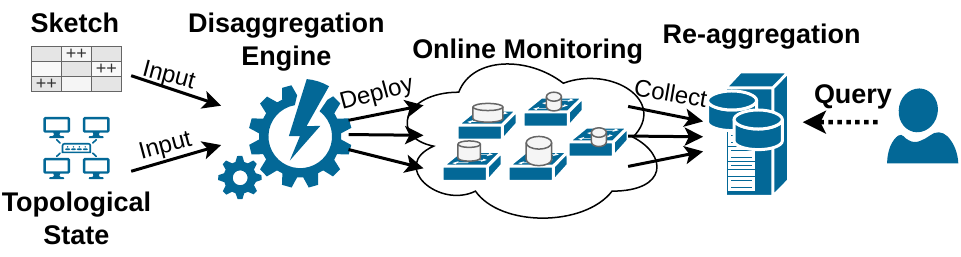}
    \vspace{-0.2in}
    \caption{Overview of Sketch Disaggregation.}
    \label{fig:disaggregation_overview}
    \vspace{-0.1in}
\end{figure}

\sloppy
To reduce costs and improve performance, practitioners offload functionalities to network devices, such as programmable network switches~\cite{nvidia-deepstream,broadcom-streaming}.
A fundamental challenge is that these have little available memory and limited computation capabilities~\cite{trident,tomahawk,tofino}. The total amount of memory available for line-rate statefulness and packet buffering is as little as O(10MB)~\cite{tofino}.
Further, these resources need to be shared across different applications, including streaming analytics~\cite{univmon,HashPipe,PINT}, security~\cite{xing2021ripple,liu2021jaqen,xing2022bedrock}, machine learning aggregation~\cite{sapio2019scaling,atp-nsdi,li2024thc}, storage for database systems~\cite{distcache,lerner2019case,tirmazi2020cheetah}, and various network functionalities~\cite{miao2017silkroad,li2019hpcc,he2021scalable,pan2021sailfish}.
As different switches may run different applications, and their requirements change dynamically (e.g., based on the traffic or query patterns), the amount of free resources varies greatly and constantly changes, making it hard to deploy any analytics solution that requires a fixed amount of memory.

\sloppy
In this paper, we propose utilizing switches' current {\em residual resources} as `fragments' of a disaggregated data structure.
Focusing on sketches, as they are fundamental building blocks for streaming analytics, we present a principled way to disaggregate a single data structure into multiple, dynamically sized pieces that can change in real time (Figure~\ref{fig:disaggregation_overview}). 
Those fragments can be centrally collected and used together to answer queries similarly to a single data structure.

While there has been a lot of research focused on minimizing the sketch size while optimizing the size-accuracy tradeoff~\cite{CMSketch,CountSketch,univmon,monterubbiano2023lightweight,chen2021precise}, these measure flow statistics at a single node, and their accuracy is thus restricted by the resources available there. 
However, this overlooks the potential benefits of leveraging residual resources at other nodes.
Our key observation is that in networks, the same packet typically traverses multiple nodes (e.g., switches), allowing us to leverage the varying memory across these nodes to improve the accuracy, even when different packets pass through different switches.
Although we focus on network analytics, our principles apply to similar settings that naturally appear in distributed databases. For instance, in distributed indexes, where key searches traverse multiple nodes with different searches following different paths, the resources along the path can be effectively harnessed~\cite{sherman-btree,distributed-btree}.

More concretely, each node fragment provides an estimate, and our methods combine these to produce a single, accurate, estimate from the ensemble along a flow's path.  
While this is conceptually straightforward, several challenges arise due to inequalities in sketch accuracy caused by heterogeneity: 
(1) nodes vary in resource availability, 
(2) nodes experience different traffic volumes,
and (3) flows traverse paths of different lengths, resulting in varying numbers of fragments forming estimations.
Furthermore, we must address switches' highly restricted computational capabilities, where even basic operations like multiplication and division may be unsupported in the hardware, and minimal computation can be performed per packet due to fast line rates.

In our solution, we consider the time divided into consecutive `epochs' that provide the granularity at which the user can express its queries.
The main innovation is to make epochs divisible, consisting of smaller `subepochs' after which the fragment is collected.
Each flow is measured during a single subepoch per epoch, allowing us to utilize the available resources at a switch more efficiently. 
This approach trades memory for reporting frequency: when memory is constrained, more subepochs are used to ensure the error is within acceptable bounds; when memory is abundant, fewer subepochs are needed, thereby reducing the collection frequency.

To answer queries, we combine the fragments that measure the flow across the nodes along its path. 
Estimates from fragments covering shorter subepochs are scaled proportionally to ensure comparability across measurements. 
The underlying assumption is that, since epochs are short, a flow's rate remains relatively uniform within an epoch, allowing accurate estimations to be inferred from a single subepoch.

We realize this technique in DiSketch -- a system that efficiently disaggregates sketches to leverage all residual resources while optimizing accuracy. 
We demonstrate the generality of DiSketch, we apply it to three popular sketches: Count Sketch~\cite{CountSketch}, Count-Min Sketch\cite{CMSketch}, and UnivMon\cite{univmon}.
We compare DiSketch against traditional sketch deployments (which we refer to as ``aggregated'') as well as versus DISCO~\cite{bruschi2020discovering}, a recent sketch disaggregation technique.
Extensive experiments show that our solution is highly memory-efficient, achieving comparable accuracy to DISCO while using only $25\%$ of the memory, or reducing error by nearly an order of magnitude under the same memory constraints.

\section{Motivation}
\label{sec:motivation}
    \begin{table}
        \centering
        \resizebox{\linewidth}{!}{%
            \begin{tabular}[b]{@{}llll@{}}
                \toprule
                \textbf{Application} & \textbf{Examples} & \textbf{Memory} \\ \midrule
                Basic Packet Processing & switch.p4~\cite{molero2022fast} & 30\% \\ \midrule

                Security & Ripple~\cite{xing2021ripple}, Jaqen~\cite{liu2021jaqen}, Bedrock~\cite{xing2022bedrock} & +10-50\%\\
                Machine Learning & SwitchML~\cite{sapio2019scaling}, ATP~\cite{atp-nsdi}, THC~\cite{li2024thc} & +10-40\% \\
                Storage/Database & DistCache~\cite{distcache}, NETACCEL~\cite{lerner2019case}, Cheetah~\cite{tirmazi2020cheetah} & +20-30\% \\ 
                Networking & SilkRoad~\cite{miao2017silkroad}, HPCC~\cite{li2019hpcc}, SwRL~\cite{he2021scalable}, Sailfish~\cite{pan2021sailfish} & +5-40\% \\ 
                
                \bottomrule
            \end{tabular}%
        }
        \caption{The on-switch memory cost of network functions.}
        \label{table:sram:applications}
        \vspace{-0.3in}
    \end{table}

    Recent advancements in streaming analytics have led to the development of compact sketch-based solutions, as described in several works~\cite{zhao2021lightguardian,univmon,SketchVisor,monterubbiano2023lightweight,huang2021toward,sketchlearn,wellem2019flexible,dong2024simisketch}.
    These studies have demonstrated the feasibility of implementing these structures within the constraints of modern switches, achieving minimal estimation errors~\cite{zhao2021lightguardian,monterubbiano2023lightweight,huang2021toward,sketchlearn}. 
    However, sketches are memory-intense data structures, and their accuracy directly depends on the amount of memory dedicated to them. As we show later, deploying sketches on switches alongside other functionality competing for limited memory can result in significant accuracy degradation.

    Our survey of recent literature on in-network computing, which encompasses applications ranging from security to machine learning acceleration and network functions, reveals a significant demand for switch resources. For instance, basic packet processing capabilities alone, such as L2/L3 forwarding, may consume up to 30\% of a switch's memory, as shown in Table~\ref{table:sram:applications}. Including additional functionalities further reduces the available memory for sketches.

\begin{figure}[]
        \centering
        \includegraphics[width=1.0\linewidth]{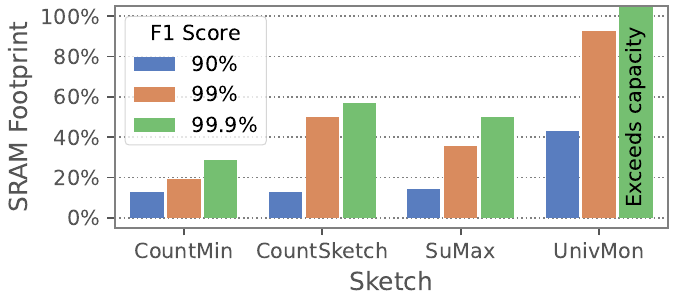}
        \vspace{-0.3in}
        \caption{On-switch memory cost to achieve an accuracy target while monitoring 30s of real-world backbone traffic.}
        \label{fig:cost_sketches}
        \vspace{-0.1in}
    \end{figure}

    \looseness=-1
    To quantify the impact of this resource competition, we analyzed the SRAM memory requirements of both established (i.e., Count Sketch~\cite{CountSketch}, Count-Min-Sketch~\cite{CMSketch}) and recently proposed sketches (i.e., UnivMon~\cite{univmon}, SuMax~\cite{zhao2021lightguardian}) for heavy hitter detection over a 30-second window (similar to previous works~\cite{elasticsketch,univmon,wellem2019flexible}), using real-world traffic traces from an Internet backbone~\cite{CAIDA2019}. 
    Our findings, depicted in Figure~\ref{fig:cost_sketches}, show that to achieve a 99\% F1 score, the memory demand of sketches ranges from 20\% to nearly 90\% of a programmable Tofino switch's SRAM~\cite{tofino}, underscoring the challenge of maintaining high monitoring accuracy while co-locating sketches with other functions.
    A high monitoring accuracy is essential as the base of responsive diagnosis~\cite{li2019deter,schlinker2019internet}, and a $99\%$ accuracy already indicates a significant amount of incorrect flow classification. Improved classification accuracy requires even more memory and can sometimes exceed the switch's memory capacity, even in isolation.
    

    Moreover, even if a sketch is not co-located with any other in-network function, it is worth noting that the amount of traffic observed by each switch significantly differs, and there can be orders of magnitude different volumes even for switches with the same logical purpose (e.g., edge switches in a datacenter)~\cite{facebook_trace,IMCtrace,delimitrou2012echo}. As a consequence, deploying the measurement on a cut in the network topology (e.g., all edge switches~\cite{harrison2018network}) results in varying degrees of accuracy even when all switches have the same memory allocated for measurement. 

    \begin{figure*}[t]
        \centering
        \begin{subfigure}[t]{0.48\linewidth}
            \centering
            \includegraphics[width=\linewidth]{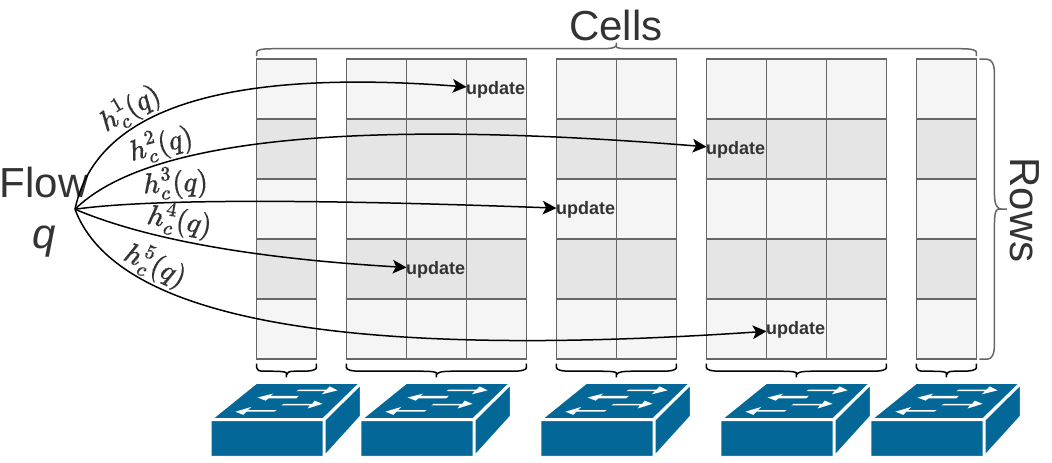}
            \caption{Per-column disaggregation. Fragments host full-depth sketches.}
            \label{fig:per_column_disagg}
        \end{subfigure}
        \begin{subfigure}[t]{0.48\linewidth}
            \centering
            \includegraphics[width=\linewidth]{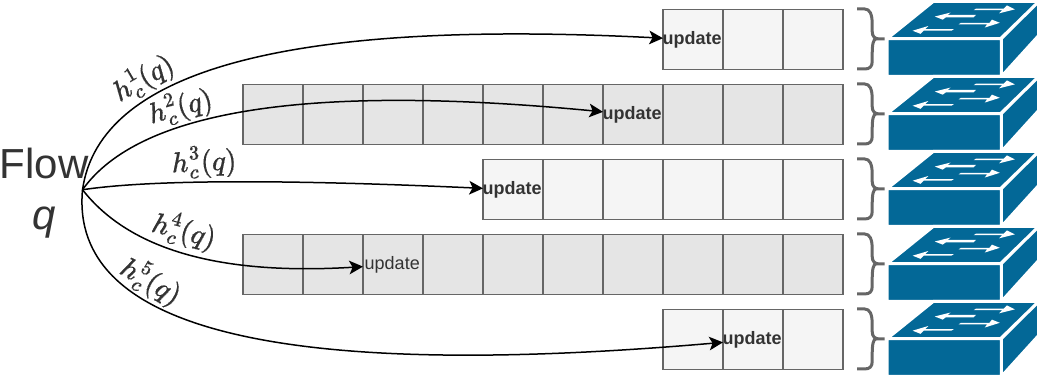}
            \caption{Per-row disaggregation. Fragments host one row each.}
            \label{fig:per_row_disagg}
        \end{subfigure}
        \vspace{-0.1in}
        \caption{Visualization of sketch disaggregation directions. Note that fragments can have different numbers of cells.}
        \label{fig:disagg_direction}
    \end{figure*}
    
    Acknowledging these challenges, disaggregating sketches across multiple switches emerges as a promising solution for utilizing network-wide resources.

\section{Sketch Disaggregation}\label{sec:sketch_disaggregation}
    We consider disaggregating \emph{sketches}, which, for our purposes, are viewed as a matrix structure in which each cell is an identical copy of a simpler data structure, usually a counter. We further assume that when an element (e.g., a packet) is inserted into the data structure, its key (e.g., flow ID) is mapped via hashes into one or more cells in each row, and the cells are updated appropriately. In what follows we assume one cell is updated in each row, and that cell is chosen uniformly in each row by the hash function.    
    Examples of such sketches include ones for frequency estimation~\cite{huang2021toward, cormode2005improved, estan2002new, CountSketch, zhao2021lightguardian}, set membership~\cite{bloom1970space, fan2014cuckoo,dong2024simisketch}, sparse recovery~\cite{ganguly2007counting,indyk2008near}, frequency moments estimation~\cite{alon1996space,univmon,chen2021out}, entropy estimation~\cite{univmon, clifford2013simple, harvey2008sketching}, and $\ell_p$ samplers~\cite{cohen2015lp,cohen2020wor}.
    These sketches' cells are exported and reset periodically, to prevent the buildup of stale data and subsequently reduced estimation accuracy. The period between these resets is referred to as the epoch.

    \begin{figure}[]
        \centering
        \includegraphics[width=1.0\linewidth]{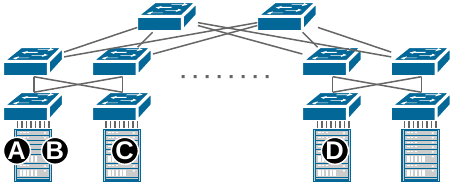}
        \vspace{-0.3in}
        \caption{A Fat-Tree Topology. Packets traverse 1 (e.g., A-B), 3 (e.g., A.C), or 5 (e.g., A-D) switches.}
        \label{fig:fattree_topology}
    \end{figure}

    Understanding the challenges of sketch disaggregation across multiple nodes begins with a depiction of a datacenter network's architecture. A classic Fat-Tree topology, commonly referenced in literature and employed in real-world deployments, is illustrated in Figure~\ref{fig:fattree_topology}. This topology highlights the existence of numerous paths between any two end-nodes, with the number of hops varying based on the nodes' locations. For example, flows between (A) and (B) traverse just a single switch, while any path between (A) and (D) contains five switches.
    Sketch disaggregation is then the process of dividing a sketch across network paths, where each network node hosts a fragment of the sketch. As in traditional deployments, we assume that at the end of each epoch, per-node data structures are sent for analysis to a central server called the controller.
    After collection and aggregation, per-path fragments can be queried together to answer queries similarly to traditional aggregated sketches.
    With this in mind, there are two natural approaches to disaggregating sketches: \emph{per-column} and \emph{per-row} disaggregation.
    
    In per-column disaggregation (Figure~\ref{fig:per_column_disagg}), each network hop would host all sketch rows of the sketch matrix, but only a part of the columns. Keys are still mapped to one cell in each row, selected uniformly at random by a hash function.
    This requires that both the path length and fragment widths have to be known by all sketch nodes when a packet is traversing the network, which is costly. That is because a switch needs to know the indices of columns it holds, as well as the total number of on-path columns, since the hash functions need to output a column index. 
    Previous work on sketch disaggregation attempts to solve this through lookup tables in each fragment containing entries for every network path going through a switch~\cite{gu2023distributedsketch}, posing a scaling issue for large networks.
    For example, on a $k$-ary Fat-Tree, each edge switch needs to store information about $k^3\cdot(k-1)/8$ paths (it has $k/2$ options for each of the aggregate and core switches, $k-1$ for the aggregate switch on the way down and another $k/2$ for the last edge switch). This means that even for moderately sized networks such as $k=28$, we would require information for about 74 thousand paths, leaving less memory for the sketch itself and undermining the original purpose of the disaggregation.
    
    In contrast, in per-row disaggregation (Figure~\ref{fig:per_row_disagg}), each node hosts a single row in each fragment, occupying all sketch-allocated memory.
    Each fragment is then equivalent to an independent sketch row, so they can function in isolation, and there is no need for a lookup table as with per-column aggregation.
    Note that, unlike standard sketches, the differences in row sizes due to memory lead to an ``irregular'' shaped matrix, which makes it harder to aggregate the results into a single accurate estimate.
    
    \begin{figure}[]
        \centering
        \includegraphics[width=1.0\linewidth]{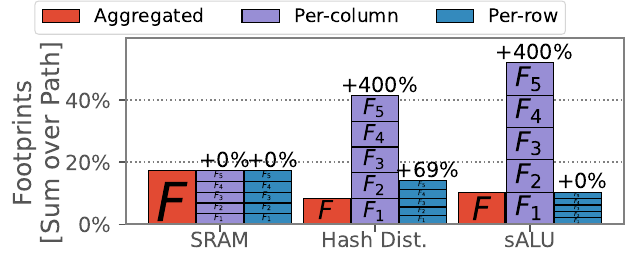}
        \vspace{-0.3in}
        \caption{The disaggregation direction has a significant impact on computational resources. Shown here are full-path footprints in a Tofino programmable switch, broken down per switch/fragment.}
        \label{fig:disaggregation_resources_comparison}
    \end{figure}
    To illustrate the disaggregation overheads, we implemented state-of-the-art per-row (DISCO~\cite{bruschi2020discovering}) and per-column (Distributed Sketch~\cite{gu2023distributedsketch}, lookup table excluded from cost)\footnote{The exclusion of the lookup table cost from Distributed Sketch emphasizes the inherently higher base footprint of per-column disaggregation, regardless of indexing technique.} disaggregated sketches on a programmable Tofino switch~\cite{tofino}.
    Figure~\ref{fig:disaggregation_resources_comparison} presents these overheads, comparing the total resources required along a path with those of a traditional Count-Sketch.

    While per-row disaggregation generally offers better resource efficiency than per-column, producing accurate estimates from this kind of data structure is particularly challenging when the switches differ significantly in the memory amount and traffic volume.  
    
    Before we describe our solution, we discuss the main challenges of disaggregation.
    
    \sloppy
    \textbf{Challenge 1:}
        Nodes across the network can have varying resources, a result of deploying distinct in-network functions at different switches and possibly other causes of switch heterogeneity. 
        Some functions, such as security mechanisms, may be more suitably deployed at switches near endpoints~\cite{xing2022bedrock,sapio2019scaling,tirmazi2020cheetah}, while others fit better within the network core~\cite{pan2021sailfish}.
        This diversity leads to a heterogeneous use of memory, ruling out per-column disaggregation due to the high computational overhead and substantial memory requirements for stateful per-flow counter allocation. 
        Per-row disaggregation in highly heterogeneous deployments can experience accuracy degradation, where tiny fragments introduce significant errors to the composite sketch. 
        Alternatively, these fragments become essentially negligible when assigned an importance proportional to their relatively high error.

    \textbf{Challenge 2:}
    \noindent
        Traffic volume can vary significantly across nodes, stemming from the design of datacenter networks and their traffic patterns.
        For example, the Fat-Tree topology facilitates massive multi-path routing~\cite{al2008scalable}, yet despite load-balancing efforts, imbalances persist~\cite{alizadeh14_conga,katta2016hula,katta2017clove,ghorbani2017drill,vanini2017let}.
        Studies have shown that much traffic remains local, with only a fraction traversing the entire datacenter~\cite{facebook_trace}. 
        Consequently, switches near end-hosts experience higher traffic volumes than those at the core, affecting the accuracy of sketch fragments under heavy loads.

    \begingroup
        \setlength{\columnsep}{6pt}%
        \begin{wrapfigure}{r}{0.4\linewidth}
            \centering
            \includegraphics[width=1.0\linewidth]{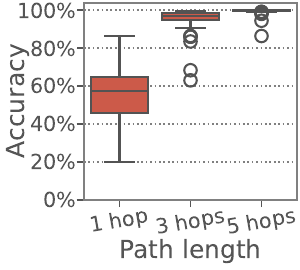}
            \vspace{-20pt}
            \caption{Path-lengths' impact on per-row disaggregation.}
            \vspace{-10pt}
            \label{fig:vertical_naive_pathlengths}
        \end{wrapfigure}        
        \textbf{Challenge 3:}
            The path lengths of different flows vary, influenced by the datacenter's network topology and traffic distribution. While some traffic remains local, affecting only a few nodes, other flows span across the network. 
            This variance means that some traffic benefits from more extensive observation by multiple fragments, whereas others do not. 
            For flows traversing only a single fragment, the accuracy degradation is akin to using a one-row sketch, which can yield significant inaccuracy. 
            To demonstrate this effect, we deploy DISCO, a per-row disaggregated count sketch, in a Fat-Tree topology using the same experimental parameters as further down in Section~\ref{sec:eval_disketch_comparison}.
            We show a breakdown of the per-path-length's impact on heavy hitter detection in Figure~\ref{fig:vertical_naive_pathlengths}.
            Notice the significant impact that the path length has on the estimation accuracy, with queries regarding single-hop flows being highly inaccurate.

        In the following section, we introduce a general technique that can be used to deploy per-row disaggregated versions of sketches to achieve highly accurate estimations. 
        
    \endgroup

\section{Spatiotemporal Disaggregation}\label{sec:spatiotemporal_disaggregation}
    Here, we introduce our spatiotemporal disaggregation solution for resource-scarce sketch disaggregation.

    In heterogeneous network environments, where nodes differ in resource capacity and traffic load, sketch accuracy can be significantly compromised due to smaller and/or overloaded fragments that introduce substantial estimation errors. In some cases, it may be more effective to entirely disregard these less reliable fragments and instead rely on the fewer, larger fragments along the network path. This is contrary to our goal of utilizing resource gaps and hints at a deeper issue with inefficient resource leverage.

    An appealing solution is flow-level sampling, where fragments track a subset of flows whose size is determined by the fragment's memory and expected traffic load. 
    Larger or less loaded fragments can handle more flows, while `congested' fragments require more restrictive sampling. 
    This approach allows even small fragments to provide useful insights on the flows that they track while larger fragments provide a more complete flow coverage.
    However, naive sampling risks leaving some flows untracked when all their on-path fragments fail to sample them. 
    Additionally, it introduces inconsistent measurement inaccuracies, with some flows favored over others even when they traverse the same path.
    This inconsistency can undermine the overall reliability of the sketch.

    To mitigate this unreliability, we propose a novel temporal sampling technique where a subset of flows are actively monitored at a time during dynamically defined periods. This way, we optimize resource utilization and accuracy while ensuring each fragment provides equal coverage for all flows.
    
    \begin{figure}[]
        \centering
        \includegraphics[width=1.0\linewidth]{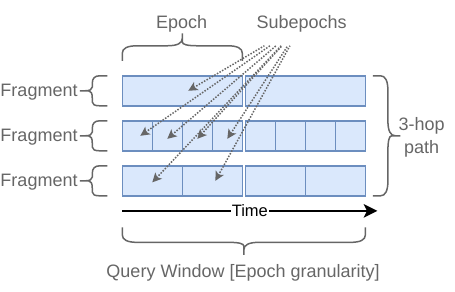}
        \vspace{-10pt}
        \caption{Sketching epochs are divided into subepochs. Fragments are single-row sketches residing on different switches. Queries are executed against composite sketches, comprising all relevant subepoch records.}
        \label{fig:epoch_terminology}
    \end{figure}
    
    We start with a brief overview of our solution, followed by a more detailed description in subsequent sections. Please refer to Figure~\ref{fig:epoch_terminology} for a visual overview of our terminology.
    \paragraph{Subepoching (\S\ref{sec:design_subepoching}): }
        Each fragment divides its sketching epoch into $n$ subepochs, where $n$ is chosen per fragment.
        A flow is only monitored during one subepoch in each on-path fragment, hence a fragment only tracks approximately $\frac{1}{n}$ of the flows at a time.
        Counters are exported and reset after each subepoch to allow for central querying.
        We elaborate on subepoching in Section~\ref{sec:design_subepoching}.

    \paragraph{Error Equalization (\S\ref{sec:design_error_equalization}): }
        A network-wide error target is chosen, which is used to homogenize errors across fragments to facilitate network-wide querying. 
        Fragments accomplish this by estimating their local errors at the end of each epoch and using this estimation to adjust the number of subepochs for the upcoming epoch. This way, all fragments aim to deliver similar error bounds in the upcoming epoch.
        We elaborate on error equalization in Section~\ref{sec:design_error_equalization}.
        
    \paragraph{Central Queries (\S\ref{sec:design_querying}): }
        We consider queries that analyze the traffic in one or several adjacent sketching epochs. 
        These queries are executed against a specific key (e.g., a network flow, port, or host) or the aggregate network (e.g., in entropy estimation).
        The exact supported queries depend on the capabilities of the deployed sketch that is being disaggregated. 
        At query time, collected data from relevant fragments are used to compile a virtual composite sketch that covers the queried data streams.
        We elaborate on composite sketch compilation and querying in Section~\ref{sec:design_querying}.

    \subsection{Subepoching}\label{sec:design_subepoching}
        \begin{figure}[]
            \centering
            \includegraphics[width=.8\linewidth]{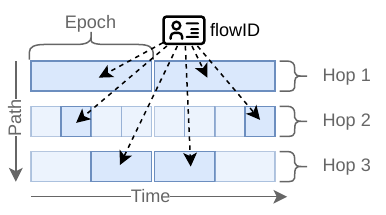}
            \caption{Flows map into one subepoch per epoch in each hop, leading to temporal sampling.}
            \label{fig:spaciotemporal_indexing}
        \end{figure}
        We now provide a detailed description of subepoching, which is the foundation of our temporal sampling technique.
        For ease of exposition, we first assume that all flows follow the same path length; in Section~\ref{sec:design_singlehop}, we discuss optimization that improves the accuracy when the path lengths differ.
        
        Sketch fragments dynamically divide their epochs into a power-of-two number of subepochs ($n$) so that $n = 2^x$ for some $x \in \mathbb{N}$.
        Specifically, during epoch $E$, each fragment $F$ divides the global epoch duration into $n^F_{E}$ equal-sized subepochs. 
        To simplify the notation, we will from now on omit the $F$ superscript from variables when there is no fragment ambiguity.
        For simplicity and efficiency of implementation, we hereafter assume that $n$ is a power of two.
        
        Subepoching allows fragments to perform temporal flow-level sampling while simultaneously guaranteeing flow queryability within each query window. 
        This is achieved by mapping each flow to one subepoch per fragment in each epoch, resulting in each subepoch monitoring a distinct subset of flows. We visualize this in Figure~\ref{fig:spaciotemporal_indexing}.
        Subepoch mapping for epoch $E$ is achieved through a fragment-specific hash function $s_E : \mathbf{q} \rightarrow \{0,1\ldots, n_E-1\}$, where $\mathbf{q}$ is the set of flows that traverse the fragment during a subepoch. That is, each flow $q$ is monitored within subepoch $s_E(q)$.
        
        A subepoch record is generated and exported at the end of each subepoch, containing all information required to centrally query the data stream (see details in Section~\ref{sec:design_querying}).
         
        While initiating a new epoch, fragments individually replace their hash functions to prevent persistent collisions.
        Additionally, the number of subepochs $n$ is recomputed based on fragments' estimated performances, to equalize the network-wide error bounds.

    \subsection{Error Equalization}\label{sec:design_error_equalization}
        In most cases, sketches provide accuracy guarantees based on an analysis in which a basic data structure (e.g., a sketch row in count min and count sketches) provides the desired precision with a constant probability (e.g., 3/4). Merging the estimates of independent repetition of this structure (e.g., using min in the Count-Min sketch or median in the Count Sketch) then amplifies the success probability as desired.
        For simplicity, and because we find it effective, we follow the same rationale -- we aim for different fragments to yield estimates with similar errors, allowing us to amplify the success probability through merging.

        With this in mind, we consider the Count-Min and Count Sketches as two examples. 
        The standard analysis of the Count-Min sketch looks at the expected noise that other flows impose onto the queried flow's counter. Assuming that the hash function is pairwise independent, any other flow increases the counter with probability $1/w$, i.e.,  the expected noise is bounded by $\frac{\sum_{k=1}^{\abs{\mathbf{q}}}f_k}{w}$. By Markov's inequality, this gives 
        \begin{equation}
        \Pr\left[   \hat{f}_k - f_k   \geq   4\frac{\sum_{k=1}^{\abs{\mathbf{q}}}f_k}{w}   \right] \leq \frac{1}{4} \ .
        \end{equation}
        
        Similarly, following the standard analysis of Count Sketch~\cite{CountSketch,larsen2021countsketches}, each fragment of width $w$ has a frequency estimate $\hat{f}_k$ for every flow $q_k \in \mathbf{q}$ with expectation $f_k$ and variance at most $\frac{\sum_{k=1}^{\abs{\mathbf{q}}}f_k^2}{w}$.  Using Chebyshev’s inequality, this implies that:
        \begin{equation}
        \Pr\left[   |\hat{f}_k - f_k|   \geq   2\sqrt{\frac{    \sum_{k=1}^{\abs{\mathbf{q}}} f_k^2}{{w}}}   \right] \leq \frac{1}{4} \ .
        \end{equation}

        Therefore, based on the switch's available space, we aim to size the subepochs such that the noise bound roughly matches a target quantity $\rho$, which we loosely refer to as a fragment's probabilistic error bound (PEB). The magnitude of a fragment's estimation errors is linked to $\rho$, and based on the above, we set:
        \begin{equation}\label{eq:sketch_error_bound}
            \rho = 
            \begin{cases}
                \sqrt{\frac{\sum_{k=1}^{|\mathbf{q}|} f_k^2}{w}}, & \text{if CS} \\
                \frac{\sum_{k=1}^{|\mathbf{q}|} f_k}{w}, & \text{if CMS}
            \end{cases} \ .
        \end{equation}

        To roughly equalize the error across fragments, we define a network-wide \emph{target PEB} $\rho_{\mathit{target}}$, representing the desired PEB in all fragments' subepoch records. Each switch, knowing its space constraints, then attempts to meet this target.

        For our purposes, we can think of sketch fragments as singe-row sketches.
        Given that sketches aim to \emph{estimate} the underlying frequency vector $\mathbf{f}$, we can use the fragment's counters $c_i \in \mathbf{c}$ as an approximation of the frequency vector. We can then modify Equation~\ref{eq:sketch_error_bound} to calculate an estimated subepoch PEB $\widehat\rho \approx \rho$ from the sketch counters:
        \begin{equation}\label{eq:estimated_subepoch_error_bound}
            \widehat\rho = 
            \begin{cases}
                \sqrt{\frac{\sum_{i=1}^{w} c_i^2}{w}}, & \text{if CS} \\
                \frac{\sum_{i=1}^{w} c_i}{w}, & \text{if CMS}
            \end{cases} \ .
        \end{equation}

        Let $\widehat\rho_{E,s}$ be the estimated PEB from subepoch $s$ in epoch $E$.
        An epoch's average error bound is then estimated as:
        \begin{equation}\label{eq:estimated_epoch_error_bound}
            \overline{\widehat\rho_E} = \frac{\sum_{s=0}^{n_E-1} \widehat\rho_{E,s}}{n_E} \ .
        \end{equation}

        We aim to equalize $\overline{\widehat\rho_E}$ across the network, so that $\overline{\widehat\rho_E} \approx \rho_\mathit{target}$ for all fragments and epochs.
        Recall that $n \approx \mathit{FlowSamplingRate}^{-1}$ within each subepoch, which leads to $\rho \propto \frac{1}{n}$.
        Further, we are assuming that traffic patterns are relatively stable between consecutive epochs, hence $\overline{\rho_{E+1}} \approx \overline{\widehat\rho_E}$ for each epoch $E$.
        Following these assumptions, fragments autonomously adjust $n$ to approach $\rho_\mathit{target}$:
        \begin{equation}\label{eq:recursive_n_computation}
            n_{E+1} = 
            \begin{cases}
                2n_E,& \text{if } \overline{\widehat\rho_E} > 2\rho_{\mathit{target}}\\
                \max(1,\frac{n_E}{2}),& \text{if } \overline{\widehat\rho_E} < \frac{\rho_{\mathit{target}}}{2}\\
                n_E,              & \text{otherwise}
            \end{cases} \ .
        \end{equation}
        $\rho_{-1}$ is not defined, so we initiate the first epoch with a likely suboptimal $n_0=1$.
        We recommend computing a moving $n_{E+1}$ as we do in Equation~\ref{eq:recursive_n_computation} instead of independently calculating it (using the equation \mbox{$n_{E+1} = 2^{\nint{\max\left(0, \log_2\frac{\overline{\hat{\rho}_E}}{\rho_\mathit{target}}\right)}}$}) to reduce the effect of $\widehat\rho$ outliers. Calculating a moving $n$ showed a slight empirical advantage in our simulations.

        One might be tempted to set an extremely low $\rho_\mathit{target}$ for the network, to increase the estimation accuracies.
        Unfortunately, $\rho_\mathit{target}$ only states the error bound for queries \emph{within a single subepoch}, but we want to optimize for the estimation accuracy during a query window, i.e., a set of contiguous epochs. The epoch estimation is, in cases of temporal blind spots, an extrapolation from the mean subepoch estimation and is therefore highly dependent on how representative the tracked flow pattern is to the full epoch. 
        Therefore, we want to minimize the final estimation error $\epsilon = \epsilon_\mathit{subepoch} + \epsilon_\mathit{extrapolation} \approx \rho + \epsilon_\mathit{extrapolation}$.
        Unfortunately, $n \approxprop \frac{1}{\rho_\mathit{target}}$, i.e., decreasing $\rho_{target}$ generally increases fragments' $n$, leading to shorter flow-tracking time windows and subsequently increased extrapolation errors $\epsilon_\mathit{extrapolation}$.
        
        The extrapolation accuracy is a function of $n$ and the distribution of inter-packet arrival times for flows (i.e., flow burstiness patterns). For instance, the frequency estimations for highly bursty flows are unlikely to be accurately estimated from a sampled time window, while uniformly transmitting flows can be accurately extrapolated.
        $\rho_\mathit{target}$ should be selected following an analysis of a network's typical traffic patterns. 

        This error equalization results in all fragments' subepoch records having similar expected error bounds, regardless of size or traffic load, facilitating accurate queries.

        \paragraph{The UnivMon Sketch:} UnivMon has a data structure that consist of multiple count sketches called `levels'. According to the above, we set each fragment to contain all levels, all of them with the same width (as in the paper~\cite{univmon}) and with the same subepoch hash.

    \subsection{Central Querying}\label{sec:design_querying}
    \ran{As a high level suggestion, I think we should separate flow queries and global (e.g., entropy) queries in the writing.}
        Each fragment exports a subepoch record $R$ at the end of each subepoch, which is sent for central collection.

        A record is defined as $R = (F, E, S, n, \mathbf{c}, \mathbf{h})$ where:
        \begin{itemize}
            \item $F$ is the fragment that the record is from,
            \item $E$ is the epoch number,
            \item $S$ is the subepoch number,
            \item $n$ is the number of subepochs that $F$ used in $E$,
            \item $\mathbf{c}$ are the counters from the end of the subepoch,
            \item $\mathbf{h}$ are the hash functions used during sketching.
        \end{itemize}
        To refer to a field, e.g., $F$, inside of a record $R$, we use $R.F$.
        Each of these records $R$ is added to the set of collected records $\mathbf{R}$ which forms the basis for network-wide queries. 

        \ran{We're defining here a query through example. Is there anywhere we list the possible queries? Also, there are no keys for entropy, which may be worth mentioning.}

        \jl{I commented out the following lines for a just-in-case early submission}
        
        %
        %
        %
        A key-based query is defined as $Q = (\theta, \tau, \kappa)$ where:
        \begin{itemize}
            \item $\theta$ is the query type (e.g., flow frequency or data volume),
            \item $\tau = [T_{\textit{start}}, T_{\textit{end}})$ is the query time window,
            \item $\kappa$ is the queried key (e.g., a flow).
        \end{itemize}
        The set of supported query types and keys depends on the capabilities of the deployed sketches.
        
        Queried time windows are at the granularity of sketching epochs so that $Q.\tau$ aligns with a set of contiguous epochs $\mathbf{E}_Q$.
        Each epoch is queried individually to yield a list of per-epoch query outputs $O_E \in \mathbf{O}$ for each $E \in \mathbf{E}_Q$. 
        These per-epoch outputs are merged to craft a final (per-key) query output $O_Q$, for instance through $O_Q = \mathit{Sum(\textbf{O})}$ or $O_Q = \mathit{Average(\textbf{O})}$ depending on the query.
            
        \vspace{.8em}\noindent
        \textbf{Each epoch $E$ is queried as follows:}
        \vspace{-.5em}


        \paragraph{Step 1 - Retrieve relevant records:}
            Let $\mathbf{R}^E_Q$ be the set of subepoch records that form the base for query $Q$ in epoch $E$.
            The specific set depends on the query, but all records have $E$ as their epoch number.
            
            Many queries, such as queries against a specific flow $Q.\kappa$ require knowledge of the network path of the flow $\mathbf{P}_\kappa$. 
            We assume that the path for flows is known or computable. This assumption is standard and can be achieved, e.g., if ECMP-based (hash-based) load balancing is used since we can recompute the hashes~\cite{gangidi2024rdma}.
            This path is used to restrict $\mathbf{R}^E_Q$ so that $R.F \in \mathbf{P}_\kappa$.

            \looseness=-1
            Further, only records from subepochs that sampled $\kappa$ are selected.
            This is done through the key-to-subepoch mapping hash function $s^F_E(\kappa)$, which is computed for each fragment $F \in \mathbf{P}_\kappa$, so that $R.S = s^F_E(\kappa)$. 
            
            We now have a set $\mathbf{R}^E_Q$ containing all records within $E$ that form the basis for query $Q$.

        \begin{figure}[]
            \centering
            \includegraphics[width=1.0\linewidth]{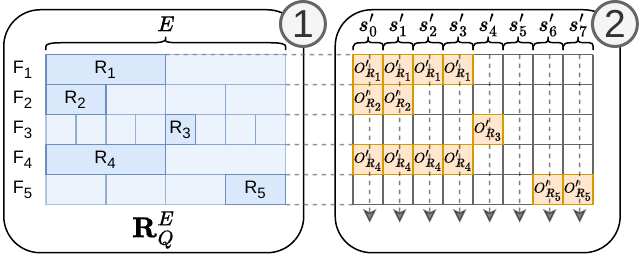}
            \caption{Centralized querying process. (1) Relevant records from on-path fragments are retrieved, with subepochs that sampled the target key during the query window selected. (2) Subepochs and estimations are normalized and merged to generate the query result.}
            \label{fig:epoch_querying}
        \end{figure}
            
        \paragraph{Step 2 - Query the Records:}
        \looseness=-1
            All records are normalized into equal-length subepochs to facilitate querying of varying-length subepoch records.
            For this, we find $n_m$, the largest $n$ for any record in $\mathbf{R}^E_Q$, which is the number of subepochs that records should normalize into.
            
            Each record $R \in \mathbf{R}^E_Q$ is queried individually as single-row sketches to retrieve their raw estimations $O_R$.
            For instance, in CMS, $O_R = c_i \in R.\mathbf{c}$, where $i = h_c(\kappa)$ and $h_c \in R.\mathbf{h}$ is the indexing hash function used at sketching-time.
            These estimations are divided into $N_R = \frac{n_m}{R.n}$ smaller estimations $O'_R = \frac{O_R}{N_R}$, one per normalized subepoch that overlap with the record's subepoch.
            
            To estimate the frequency for a key (e.g., flow) within an epoch, we first estimate its statistic within each normalized subepoch.
            Namely, for a given normalized subepoch $S'$, we consider all records the key was mapped to that include $S'$.
            
            For example, consider a case where the queried key traversed five fragments and is recorded in subepoch records $R_1,\ldots,R_5$ as shown in Figure~\ref{fig:epoch_querying} Step 1. In particular, this means that $n_m=8$ and we thus have eight normalized subepochs. 
            The estimate in the first two normalized subepochs is then based on the normalized estimations $O'_{R_1}, O'_{R_2}, O'_{R_4}$, the third is based on just $O'_{R_1}$ and $O'_{R_4}$, etc. The method for combining the estimates from each normalized subepoch depends on the sketch itself (e.g., the minimum of their estimates in Count-Min, or the median in Count Sketch).

            We note that there can be some normalized subepochs without any estimates (that is, a temporal `blind spot' for this key). 
            In this case, we use the mean of the estimates of the other normalized subepochs.
            For example, in Figure~\ref{fig:epoch_querying} Step 2, since we have no measurements of the key size during the sixth normalized subepoch, we use the average of the other seven normalized subepochs' estimates.
            These temporal blind spots are the cause of extrapolation errors when the mean is a poor estimator for the key's true statistic during this time window.

            Finally, the sum of the normalized subepoch estimates serves as a cumulative estimate over the entire epoch and is the epoch's output.

    \jl{Uncomment the empty section below}
    
    \subsection{Enhanced Single-hop Sketching}\label{sec:design_singlehop}
        As previously discussed in Section~\ref{sec:sketch_disaggregation} and shown in Figure~\ref{fig:vertical_naive_pathlengths}, the accuracy of flow estimations is closely tied to the number of fragments traversed (i.e., to the path length).
        Flows that traverse a single fragment, notably, have much worse performance, as they do not obtain the benefit from multiple estimates, and this is especially damaging when that fragment contains relatively few counters.
        We, therefore, offer a mitigation strategy that aims to enhance the accuracy of flows that traverse just a single network hop, with only mild costs in accuracy for other flows.
        For this, we assume that fragments can identify single-hop flows at sketching-time\footnote{Single-hop flows can be identified, for example, when a switch is neither receiving nor sending a packet to another switch, in cases where all network switches contain a sketching fragment.}.
        
        We arrange that these flows are monitored not in one, but two subepochs within each epoch.
        This is done by modifying the key-to-subepoch mapping to compute the two subepochs $S_{E,0} = s_E(\kappa)$ and \mbox{$S_{E,1} = s_E(q) + \frac{n_E}{2} \bmod n_E$.}
        Thus, for $n_E \geq 2$, single-hop flows are monitored during two subepochs, one in the first half and one in the second half of the epoch. If $n_E = 1$, then this mitigation technique is not applicable but also suggests that the single-hop fragment is large enough to deliver suitably accurate estimations on its own.
        Note that this technique requires a query to use two subepoch records from each queried epoch for such flows.

        Although this mitigation technique enhances the accuracy of single-hop flows, it simultaneously imposes (approximately) twice as many counter increments for those same flows, which increases the fragment's $\rho$. Therefore, we expect slightly increased errors in queries against other flows. We evaluate this effect in Section~\ref{sec:eval_pathlen}.


    \section{Implementation}\label{sec:implementation}
        \begin{figure}[]
            \centering
            \includegraphics[width=1.0\linewidth]{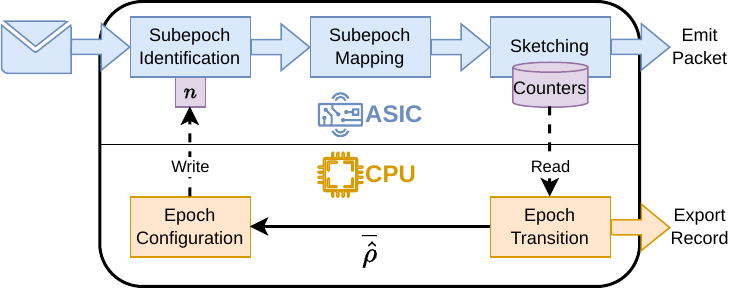}
            \caption{An in-hardware breakdown of spatiotemporal disaggregation. The high-speed ASIC does online sketching, while the slower on-switch CPU handles epoch transitions.}
            \label{fig:implementation}
        \end{figure}
        Spatiotemporal disaggregation is designed to be hardware-friendly by adhering to the restrictions of PISA (Protocol Independent Switch Architecture)\cite{bosshart2013forwarding,p4}, and can therefore process all traffic at terabit per second line-rate in current-generation switching hardware. 
        To achieve this, the online sketching components of our solution are based purely on integer arithmetic, assumed hardware-supported hash functions (e.g., CRC), and exact-match table lookups.
        However, epoch configuration requires more complex computations (see Sections~\ref{sec:design_subepoching}~\&~\ref{sec:design_error_equalization}). Due to this, we have designed the solution to only require these computations during epoch transitions, and therefore only require these at the timescales of epochs. Thanks to this, we can place these functions at a slower ASIC-adjacent CPU, as shown in Figure~\ref{fig:implementation}, a standard high-speed switch component to perform typical switch control and management. 
        
        The method for resetting the counters after each epoch is not covered in this paper. However, a possible solution is to deploy two sketches in parallel, where one sketch is actively collected and reset, while the other sketch is actively sketching~\cite{zhao2021lightguardian}. However, this is a highly inefficient use of memory, and we propose an alternative solution where the counters are \emph{not} reset each epoch. Instead, we suggest that querying and $\rho$-computation is based on the counter delta from the previous subepoch.
        
        The most computationally intensive operation that we impose on the ASIC is subepoch identification and mapping.
        For this, fragments need to identify the current subepoch to decide if a flow should be monitored.
        We propose two different solutions for identifying the current subepoch:

        \paragraph{Method 1 - Direct:}
            Each fragment could use a subepoch identification counter that increments at the end of each subepoch and resets to $0$ at the end of each epoch.
            While this approach is straightforward, it imposes unnecessary overhead in the PISA hardware architecture. It requires a per-packet conditional to detect when to trigger the incrementation, along with dedicated memory logic for the subepoch identification counter. 

        \paragraph{Method 2 - Time-inferred:}
            \begin{figure}[]
                \centering
                \includegraphics[width=1.0\linewidth]{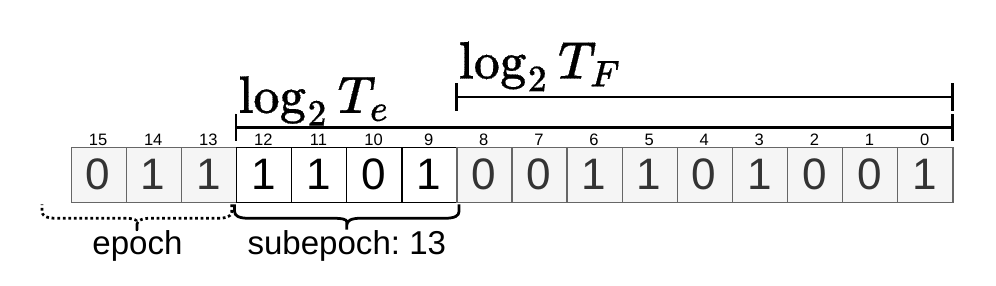}
                \caption{Infering the subepoch from a timestamp.}
                \label{fig:subepoch_bitstring}
            \end{figure}
            The current subepoch can be determined directly from a timestamp $T$ by limiting the global epoch duration ($T_e$) to a power-of-two number of time units and starting the epoch when the global timestamp is a multiple of $T_e$.
            For a fragment $F$ with $n$ subepochs, the subepoch duration $T_F = \frac{T_e}{n}$.
            The current subepoch number can then be retrieved from the timestamp bitstring between bit positions $\log_2(T_e)$ and $\log_2(T_F)$.
            For example, as shown in Figure~\ref{fig:subepoch_bitstring}, a fragment with $n=16$ knows it is currently in subepoch $13$ by extracting the substring $T[12:9] = 0b1101$, where \(\log_2(T_e)=13\) and \(\log_2(T_f) = 9\).
            This approach reduces the hardware resource footprint, at the cost of restricting epochs to power-of-two durations and offsets.

\begin{figure*}{}
    \centering
    \begin{subfigure}{\linewidth}
        \centering
        \includegraphics[width=\linewidth]{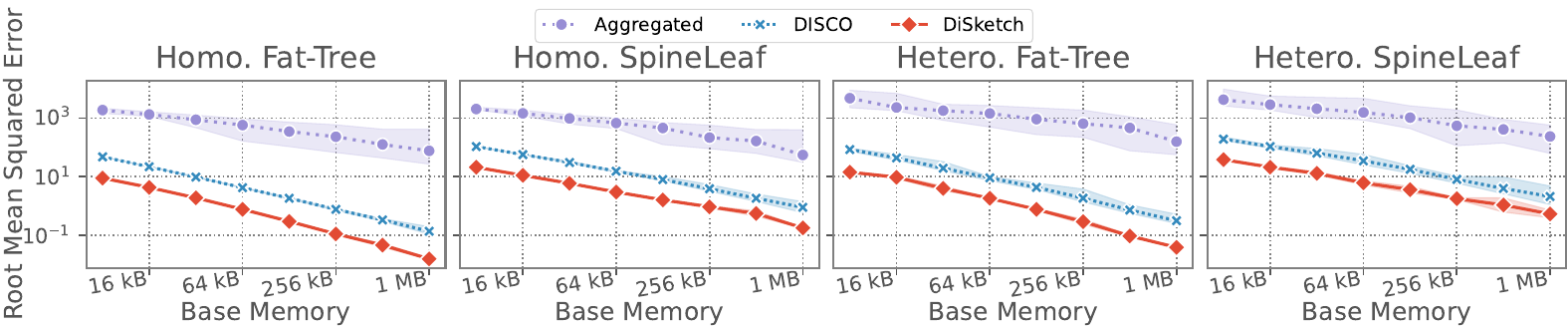}
        \vspace{-6mm}
        \caption{Evaluation results for a Count Sketch.}
        \label{fig:resgap_cs}
    \end{subfigure}
    \begin{subfigure}{\linewidth}
        \centering
        \includegraphics[width=\linewidth]{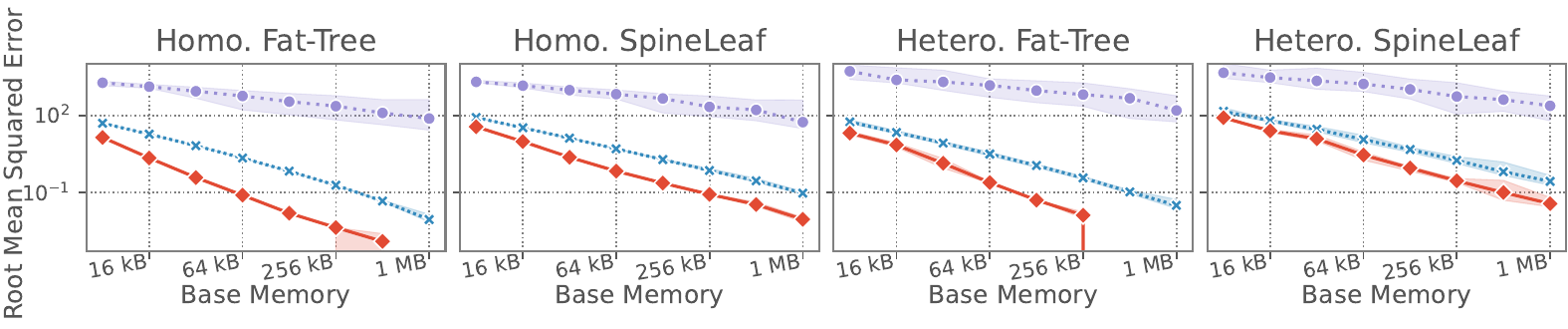}
        \vspace{-6mm}
        \caption{Evaluation results for a Count-Min Sketch.}
        \label{fig:resgap_cms}
    \end{subfigure}
    \begin{subfigure}{\linewidth}
        \centering
        \includegraphics[width=\linewidth]{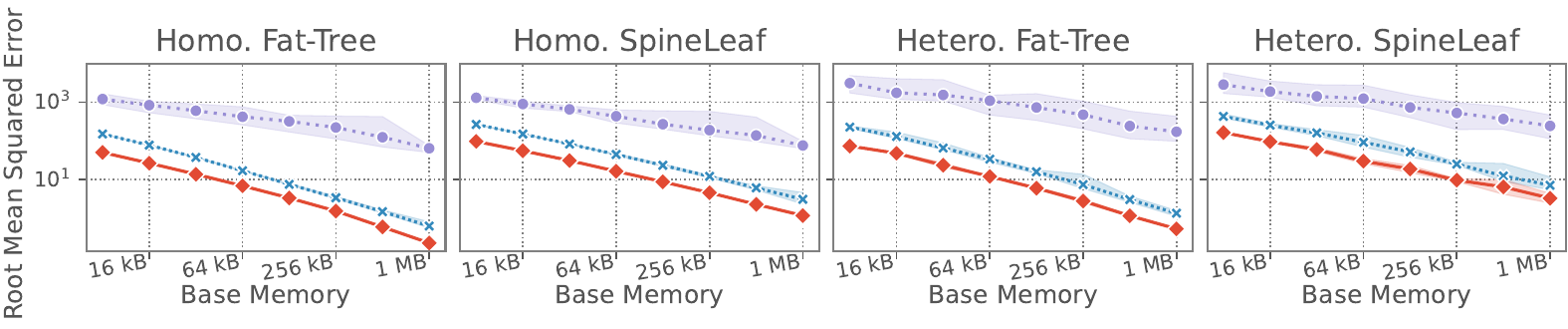}
        \vspace{-6mm}
        \caption{Evaluation results for UnivMon.}
        \label{fig:resgap_um}
    \end{subfigure}
    \caption{Frequency estimation errors of sketches for 5-hop flows at various memory sizes. Scenarios are combinations of the Fat-Tree and SpineLeaf topologies, either with homogeneous or heterogeneous switch memory sizes.}
    \label{fig:eval_resgap_accuracy}
\end{figure*}
    
\section{Evaluation}\label{sec:evaluation}
    \textbf{Topologies.}
        We simulate two smaller data center topologies: a $k=2$ Fat-Tree network with four core switches (20 switches in total), and a SpineLeaf network (12 switches in total)\footnote{The simulated networks are humble in size due to time restrictions during simulations. For context, a single data point in our experiments already requires nearly a full hour of simulation time. We have simulated topologies up to approximately double these sizes, without any noticeable impact on the results.}.

        In these networks, we evaluate two distribution scenarios for available switch memory sizes: \emph{Homogeneous} and \emph{Heterogeneous}. \\
        \emph{The heterogeneous scenario} allocates pseudo-random memory sizes for per-switch sketches. These memories are generated with a pre-defined average memory size and inter-switch heterogeneity, designed to correspond with a selected heterogeneity level\footnote{We used the gini inequality index while generating the random memory distributions.}. 
        The generated memories are randomly distributed to switches, not considering their topological position.
        As a default, we use an arbitrary heterogeneity level of $\text{gini}=0.4$, resulting in notable memory size variation between switches.
        An example memory size distribution for five switches with $\text{gini}=0.4$ can be: [$10\%$, $30\%$, $100\%$, $160\%$, $200\%$], i.e., fragment-sizes ranging from $10\%$ to $200\%$ of the base memory size.\\
        \emph{The homogeneous scenario} allocates exactly the base memory size for all network switches, resulting in a heterogeneity level of $0$.
    
    \textbf{Sketches.}
        We are evaluating spatiotemporally disaggregated instances of Count Sketch (CS), Count-Min Sketch (CMS), and UnivMon (UM)\footnote{We deploy UnivMon with 16 levels, which is approximately the second logarithm of the expected number of flows.}. 
        To simplify the discussion, we name applied versions of our solution as \emph{DiSketches} (Disaggregated Sketches), i.e., DiSketch-CS, DiSketch-CMS, and DiSketch-UM.
        For comparison, we also evaluate aggregated sketches on core switches in the network, and DISCO~\cite{bruschi2020discovering} deployed network-wide. All sketches/fragments utilize all available memory of their respective switches.

    \textbf{Traces.}
        In this evaluation, we use the real-world CAIDA-NYC Equinix packet trace \cite{CAIDA2019}, recorded at a backbone link in 2019.
        If nothing else is stated, then we replay $\sim$5 seconds of traffic ($\sim$2M packets, covering  $\sim$200K flows).
        We only have access to traffic recorded at a single network link, and we map IP addresses in traffic uniformly at random to hosts in our network to simulate network-wide communication, while omitting flows where both the source and destination map to the same host.
        
        DiSketch converges on an optimal epoch configuration over time, and we split the query window into arbitrary $32$ epochs to allow DiSketch to converge. We do, however, include the suboptimal initial epochs in the query window.

    \subsection{Frequency Estimation Errors}\label{sec:eval_disketch_comparison}
        In Figure~\ref{fig:eval_resgap_accuracy}, we present the accuracy of flow frequency estimation, a common sketch query~\cite{CountSketch,CMSketch,univmon}.
        Here, we query all flows that were tracked during the experiment in terms of their number of transmitted packets and define the query error as the absolute difference from the ground truth number of packets.
        We use the aforementioned experimental setup and replay traffic across the entire network topology. In this figure, only flows with full-length network paths are queried, to ensure that all evaluated flows would have traversed an aggregated sketch residing on core switches. We evaluate per-path-length performance later in Section~\ref{sec:eval_pathlen}.

        This evaluation demonstrates a clear pattern, where our solution consistently outperforms both aggregated deployments and DISCO-disaggregated deployments of all three example sketches. This pattern holds for homogeneous and heterogeneous deployments in both of the evaluated network topologies.
        As an example, in a heterogeneous Fat-Tree deployment, DiSketch-CS delivers $\mathit{RMSE} < 1$ using on average 128KB of per-switch memory, which is approximately $25\%$ of the 512KB that DISCO requires. 
        All aggregated sketches failed to deliver this accuracy under the evaluated memory constraints.
        Alternatively, using the same 512KB of average per-switch memory, the aggregated CS, DISCO-CS, and DiSketch-CS deliver $\mathit{RMSE}$ of 460, 0.8, and 0.09 respectively.
        These patterns hold when the sketching epoch, or the underlying traffic volumes, increase.
        We see similar patterns for all evaluated sketches and network topologies.
        This does, however, only investigate flows with full-length network paths, at fixed heterogeneity levels. We will demonstrate the effect of these parameters in sections~\ref{sec:eval_heterogeneity}~and~\ref{sec:eval_pathlen}.

        \vbox{
        \textbf{Takeaway: } DiSketch consistently and significantly enhances the memory-vs-accuracy tradeoff for frequency estimation, either increasing the accuracy by nearly an order of magnitude or reducing the required memory by approximately $75\%$ depending on the sketching environment.
        }
    \subsection{Entropy Estimation Errors}
        \begin{figure}[]
            \centering
            \includegraphics[width=\linewidth]{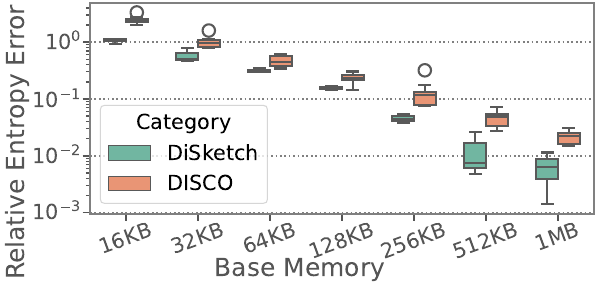}
            \caption{The error in entropy estimation of UnivMon in a Fat-Tree network with a random memory distribution.}
            \label{fig:eval_resgap_entropy}
        \end{figure}
        In Figure~\ref{fig:eval_resgap_entropy}, we present the experimental results of estimating the network-wide entropy of IP addresses. This is done through UnivMon, according to the algorithm outlined in their paper~\cite{univmon}.
        Given that disaggregated sketches far outperform aggregated sketches, we here compare DISCO only with DiSketch. We therefore also remove the full-path-length limitation, and base these evaluations on all network traffic.
        We present only the heterogeneous Fat-Tree scenario here, but the results are similar in all evaluated scenarios.
        
        Similar to the previous section, we note a consistent and significant error reduction through our solution, and DiSketch-UM delivers a similar entropy estimation accuracy as DISCO-UM using half as much memory. Alternatively, DiSketch-UM lowers the errors by approximately $50\%$ across tests.

        \vbox{
        \textbf{Takeaway: } Our solution halves the amount of memory required for entropy estimation, or halves the errors while keeping the memory unchanged.
        }
        
    \begin{figure*}[]
        \centering
        \includegraphics[width=1.0\linewidth]{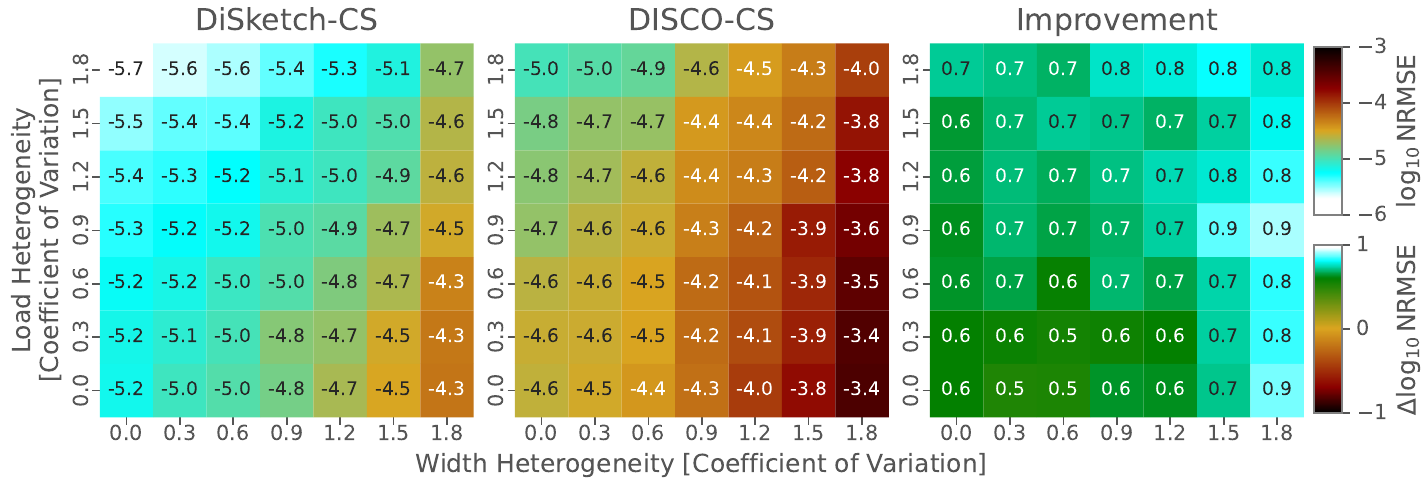}
        \caption{Heterogeneity's impact on frequency estimation, showing the $\log_{10}(\text{NRMSE})$ at various heterogeneity combinations. The right-most heatmap shows the $\log_{10}(\text{NRMSE})$ improvement over DISCO gained through spatiotemporal disaggregation.}
        \label{fig:eval_heterogeneity}
    \end{figure*}
    \begin{figure}[]
        \centering
        \includegraphics[width=1.0\linewidth]{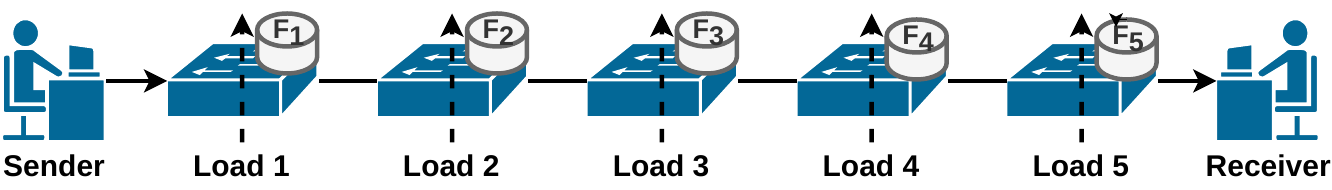}
        \caption{Experimental Setup in Heterogeneity Tests.}
        \label{fig:eval_heterogeneity_setup}
    \end{figure}
    \subsection{Heterogeneity Effects}\label{sec:eval_heterogeneity}
        Here, we evaluate DiSketch in a wide range of heterogeneous environments (i.e., where the traffic load and memory sizes vary between switches). 
        For simplicity, we again focus on the frequency estimation accuracy of DISCO-CS and DiSketch-CS.

        \textbf{Experimental Setup.}
            As opposed to the prior experiment, this experiment simulates a single 5-hop network path as shown in Figure~\ref{fig:eval_heterogeneity_setup}.
            The network load heterogeneity depends on various factors, including the network topology, load balancing, node positions, and the traffic patterns of connected hosts.
            By simulating a single path, we gain precise control of the per-switch traffic volumes and can set the heterogeneity levels freely. Traffic on other semi-overlapping paths is emulated as per-switch background traffic.
    
            We vary the memory and traffic load heterogeneity of the hops, according to the coefficient of variation (CoV, i.e., the relative standard deviation). 
            Heterogeneity levels range from perfectly homogeneous ($\text{CoV} = 0$) to significantly heterogeneous ($\text{CoV} = 1.8$).
            The total amount of background traffic ($259K$ packets) and the number of on-path counters ($5120$) is fixed across tests\footnote{The small scale in these simulations allowed us to evaluate numerous heterogeneity settings within a reasonable time, and yields the same general pattern as full-scale heterogeneity simulations.}.
            Two pseudo-random lists of 5 integers are generated for each test, one with the per-hop load, and the other with the per-fragment memory size, so that both lists adhere to the experimental parameters. 
            For instance, an example background traffic distribution with $\text{CoV}\approx1.5$ is [$204189(78.7\%)$, $18364(7.1\%)$, $29(0.01\%)$, $2265(0.9\%)$, $34675(13.4\%)$].
            The generated loads and widths are independent and are randomly distributed to the simulated nodes.
    
            There are six different packet streams: five background traffic streams, each passing through one switch, and one evaluation stream that traverses the entire path.
            The aforementioned CAIDA packet trace is used as the basis of all network traffic, and we map IP addresses at random into the traffic streams. 
            Packets in each stream are replayed chronologically in the order they appear in the packet trace.
            Approximately 300 flows comprising approximately $1\%$ of the total traffic are replayed across the full path, from sender to receiver, and are used for evaluation. The remaining $99\%$ of traffic only \emph{crosses} the individual nodes (see Load in Figure~\ref{fig:eval_heterogeneity_setup}). The experimental results are only based on flows traversing all five of these fragments.
    
            The experimental results are provided in terms of the Normalized Root Mean Squared Error (NRMSE)~\cite{basat2021salsa} between the estimated frequencies and the known ground truth, and is normalized by the total number of packets to provide a dimensionless measure of error.
            Smaller values indicate better performance.
            
        \textbf{Results.}
            The average results following multiple simulations are presented in Figure~\ref{fig:eval_heterogeneity} for all heterogeneity combinations.
            To further help illustrate the effect of DiSketch, we present the difference in NRMSE between DiSketch and DISCO in the rightmost heatmap.

            There are several patterns in this data, including an apparent direct link between the width/load heterogeneity and the sketching accuracy.
            The width heterogeneity pattern appears valid, in that we expect increased heterogeneity in per-switch memory to reduce the overall sketching accuracy in the network.
            This effect is intuitively explained by the theory behind sketches, in that a single row on its own might be inaccurate, but combining several rows boosts the accuracy super-linearly. A high width heterogeneity means that a few switches have most of the memory, leading to fewer suitably accurate rows, degrading performance when the total memory remains fixed.
            However, the apparent beneficial effect of load heterogeneity on the sketching accuracy is a likely experimental artifact. Recall that we are only evaluating the accuracy of flows traversing the full path, and the total amount of background traffic remains constant. Therefore, as we increase the load heterogeneity of the background traffic, we condense it into fewer switches, leading to an accuracy increase for most on-path fragments.
            This effect might not hold if those background flows were accounted for in the evaluation.

            We highlight the right-most heatmap, which shows the accuracy improvement as you replace DISCO with DiSketch.
            There is a consistent accuracy enhancement in every heterogeneity combination, with greater gains as the heterogeneity levels increase. This is expected, as DISCO is heterogeneity unaware, while DiSketch fragments mitigate the heterogeneity penalties by autonomously reconfiguring themselves to equalize the network-wide PEB.
            Note that the improvement is presented in terms of the absolute change in $\log_{10} \text{NRMSE}$, so a value of $1$ would signal an order of magnitude reduction in NRMSE.

        \vbox{
        \textbf{Takeaway: } 
            DiSketch outperforms DISCO in all heterogeneity combinations tested. This is especially evident as the heterogeneity levels increase, demonstrating the effectiveness of network-wide PEB equalization.
        }
        
    \subsection{Path Length Effects}\label{sec:eval_pathlen}
        \begin{figure}[]
            \centering
            \includegraphics[width=1.0\linewidth]{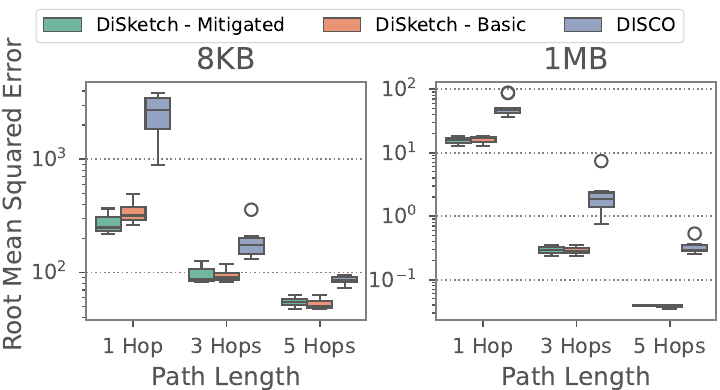}
            \caption{The path lengths' impact on frequency estimation accuracy for Count Sketch. DiSketch is evaluated both with and without the single-hop mitigation strategy.}
            \label{fig:eval_path_lengths}
        \end{figure}
        
        Here, we evaluate how the path length impacts the accuracy of DiSketch-CS and DISCO-CS.
        The heterogeneous Fat-Tree scenario from previous sections is reused, and the query results are grouped according to the path lengths of the queried flows.
        For ease of presentation, we only present the experimental results with the smallest base memory (8KB) and the largest base \mbox{memory~(1MB)~in Figure~\ref{fig:eval_path_lengths}.}

        As expected, increasing the number of traversed sketching fragments (i.e., the path length) improves query accuracy across all disaggregated sketches. 
        This effect is most significant when more memory is allocated to sketching, with single-hop queries experiencing approximately a $50$x increase in errors over 3-hop flows in the 1MB experiments, compared to a more reasonable $2.8$x increase in the 8KB experiments.
        When applying the mitigation technique from Section~\ref{sec:design_singlehop}, errors for single-hop flows decreases by approximately $24\%$ and $13\%$ in the 8KB and 1MB tests, respectively.
        Further, DiSketch demonstrates a consistent accuracy improvement over DISCO for all evaluated path lengths.
        The mitigation results in a slight error increase for queries of multi-hop flows, due to the increase in increments from single-hop flows, although the effect is within one standard deviation in these experiments.

        \textbf{Takeaway: } 
            The query accuracy is greatly impacted by the path length of the underlying flow, and the single-hop mitigation results in a slight accuracy enhancement for single-hop flows and a slight accuracy decrease for multi-hop flows. Accordingly, whether mitigation is worthwhile may depend on the setting.

\section{Discussion}
    This section provides a brief discussion of spatiotemporal disaggregation and proposes future research into the techniques presented in this paper.

    \noindent\textbf{Other data structures.} 
        This paper presented spatiotemporal disaggregation when coupled with Count Sketch, Count-Min Sketch, and UnivMon.
		However, we expect these ideas to be useful beyond that and envision that they can be applied to other sketches (e.g., Bloom filters, and HyperLogLog), as well as non-sketch data structures. 
		The exact technical requirements for spatiotemporal disaggregatability, as well as structure modifications required in each case, are left as future work.
        However, non-cumulative estimations likely have to modify the temporal merging at query time, while the spatial merging likely follows the row-merging logic of an aggregated version of the sketch, similar to how we spatially merge Count Sketch fragments through the median.
    
    
    
    \noindent\textbf{Path stability.} 
            Some load balancing schemes, such as flowlet switching~\cite{vanini2017let}, frequently alter the path of flows at incredibly short timescales. These techniques result in irregular and unstable flow paths, impacting the practicality of sketch disaggregation. 
            For instance, if flow paths change during a sketching epoch, then portions of the flow increments within that epoch could end up in different sets of fragments. Without awareness of these path changes, the estimation accuracies of the analysis engine could suffer. 
            If the analysis engine is unaware of such paths, then the estimation accuracies would suffer.
            However, it is possible to design disaggregation techniques that accommodate path changes. Assuming fine-grained path tracing is already implemented, the analysis engine could incorporate all fragments traversed during the epoch into the composite sketch output. Weighting could be employed based on the duration during which a flow has traversed each fragment.
            The development of precise techniques to perform sketch-based estimations under these conditions remains an area for future work.

            Alternatively, one could configure the load balancing techniques to only perform re-routing at epoch transitions. This adjustment would ensure that each sketching epoch contains the same set of fragments, except in cases of re-routing triggered by failures. This method could stabilize the path data within each epoch, improving the consistency and accuracy of sketch-based monitoring.

    \noindent\textbf{Finding an Optimal $\rho_\mathit{target}.$}
        Spatiotemporal disaggregation is built around subepoching, where the monitoring epoch is dynamically divided into briefer subepochs based on fragments' PEBs ($\rho$).
        However, one critical aspect has not been investigated: what is the optimal $\rho_\mathit{target}$?
        This choice is strongly influenced by the network characteristics where the sketch is deployed, including the expected loads, fragment sizes, and burstiness of flow traffic patterns.
        Determining a theoretically optimal $\rho_\mathit{target}$ is beyond the scope of this paper and is left for future work.
        However, in our experience, the selection is relatively forgiving, with any value within a factor of two of the optimal $\rho_\mathit{target}$ yielding similar performance.
        Regardless, this warrants an in-depth theoretical analysis, and experimental validation to quantify the impact of this choice across diverse network conditions.

\section{Related Work}
    \label{sed:related}

    Traditionally, sketches have been deployed aggregated~\cite{CMSketch,CountSketch,univmon}.
    Recent interest has shifted towards network-wide deployments to enhance measurement flexibility~\cite{zhao2021lightguardian,elasticsketch,bruschi2020discovering,cornacchia2021traffic,gu2023distributedsketch,li2024distributedsketchdeployment}.

    For example, Zhao et. al., 2021,~\cite{zhao2021lightguardian} designed a sketch-based network-wide telemetry system named LightGuardian.
    In their approach, each switch hosts two \emph{SuMax} sketches: one actively populated and one being collected.
    This supports new sketch measurements, including latency jitter and packet loss detection, using a probabilistic in-band collection method to reduce centralized collection costs.
    Nonetheless, this system does not address heterogeneous environments, incurs substantial resource costs, and employs aggregated (i.e., non-disaggregated) sketches on each switch.

    To our knowledge, the first paper describing the potential of disaggregated sketches was DISCO~\cite{bruschi2020discovering}, published in 2020. They argued for sketch disaggregation to simplify sketch deployments in resource-scarce environments. They presented a basic technique for per-row disaggregation of sketches for estimating flow size, showing an increased accuracy in heavy hitter detection. Although promising, they did not investigate heterogeneous environments, hardware feasibility, or more complex sketches for tasks unrelated to flow size estimation. 

    Cornacchia et al., 2021,~\cite{cornacchia2021traffic} highlighted the detrimental effects of traffic patterns on per-row disaggregated sketches, notably increased hash collisions and accuracy degradation due to load imbalances. 
    They proposed that sketch fragments sample a subset of traffic to process, but their algorithm assumes full in-band knowledge of per-flow paths and fragment dimensions, thus introducing considerable overheads and limiting deployment flexibility.
    We believe that their fundamental idea is valid, that sampling techniques would lead to efficient distributed sketching in heterogeneous environments, but that their imposed assumptions are unreasonable in many deployments. Hence, we choose not to include this solution in the evaluation. 
    DiSketch fragments operate autonomously, without any per-flow knowledge assumptions.
    We discussed the issues with these assumptions in Section~\ref{sec:sketch_disaggregation}.
    
    Gu et al., 2023,~\cite{gu2023distributedsketch} proposed per-column disaggregation as an alternative to per-row disaggregation and discussed how to handle load imbalances. 
    As outlined in Section~\ref{sec:sketch_disaggregation}, this approach faces the challenge of evenly distributing counters across diverse paths to ensure balanced counter incrementation relative to fragment sizes.
    To address this, the authors propose the use of lookup tables for counter allocation, containing one entry for every active traffic flow. 
    However, since the number of sketch cells typically grows sub-linearly to the number of keys, the inclusion of such a lookup table contradicts this fundamental design goal. Further, per-column disaggregation is inefficient in high-performance switching architectures such as PISA, since per-row computational logic remains dedicated even for packets where it is not utilized, leading to a high resource footprint.

    \label{related::DistributedSketch}
    Li et al., 2024,~\cite{li2024distributedsketchdeployment} addresses the traffic imbalance issue by proposing a deployment and increment strategy that ensures load balancing across sketch rows.
    Their method selectively deploys rows across the network, supporting a variable number of rows per fragment.
    The ingress switch determines the number of rows each hop should process per packet, inserting this information as a new header for ingressing packets and allocating traffic based on hop capacities.
    However, this places high burdens on ingress switches, requiring extensive knowledge of network paths and fragment dimensions, and reduces the MTU of the network through extended packet headers, risking frame fragmentation and reduced goodput.
    DiSketches has none of these issues.

\section{Conclusion}
    We introduced spatiotemporal sketch disaggregation, a novel sketching scheme for disaggregated streaming analysis.
    This approach combines spatial and temporal indexing through subepoching, enabling flexible and heterogeneous deployments of multiple sketches, as demonstrated with Count Sketch, Count-Min Sketch, and UnivMon.

    Our findings show that spatiotemporal disaggregation significantly enhances sketch accuracy, reducing estimation errors by nearly an order of magnitude compared to DISCO, and by several orders of magnitude over conventional aggregated sketching methods. 
    These accuracy improvements are especially pronounced in heterogeneous environments, where network-wide equalization of probabilistic error bounds allows fragments of varying loads and sizes to be efficiently queried together.
    
    \bibliographystyle{plain}
    \bibliography{references}

    \appendix

\end{document}